\documentclass[aps,prl,reprint]{revtex4-2}
\usepackage[utf8]{inputenc}
\usepackage{graphicx}
\usepackage{amsmath}
\usepackage{physics}
\usepackage{comment}
\usepackage{csquotes}

\usepackage[normalem]{ulem}
\newcommand{\stkout}[1]{\ifmmode\text{\sout{\ensuremath{#1}}}\else\sout{#1}\fi}

\usepackage{xcolor}

\begin{document}
\title{
Observation of microcavity fine structure
}
\date{\today}

\author{C. Koks}
\author{F. B. Baalbergen}
\author{M. P. van Exter}
\affiliation{Huygens-Kamerlingh Onnes Laboratory, Leiden University,
P.O. Box 9504, 2300 RA Leiden, The Netherlands}

\begin{abstract}
We observe fine structure in the resonance spectra of optical microcavities.
We identify the polarization-resolved modes in the spectrum and find that resonance frequencies split in accordance with the theoretical prediction. 
The observed fine structure is dominantly caused by an optical spin-orbit coupling and non-paraxial propagation and reflection.
Both effects are intrinsic, i.e. present in an ideal rotation-symmetric system, and scale inversely proportional to the mirror radius of curvature.
For cavities with a small radius of curvature, the influence of fine structure on the resonance spectrum is important and unavoidable and should thus be taken into account.
\end{abstract}

\maketitle

The fine structure in atomic spectra has revealed perturbations to the Hamiltonian which are intrinsic for atoms \cite{Sommerfeld1940}.
The Bohr model in atomic physics predicts groups of degenerate orbitals, labeled by the principal quantum number $n$. 
This degeneracy is lifted by perturbations such as spin-orbit coupling and a relativistic correction \cite{thorne2017}.
We observe a spectral fine structure in optical microcavities which reveals similar intrinsic perturbations to the paraxial wave equation.
The paraxial model predicts groups of frequency-degenerate transverse modes, labeled by the transverse order $N$ \cite{Siegman}.
Also their degeneracy is lifted by perturbations which are intrinsic to microcavities.

The microcavity fine structure becomes relevant when the radius of curvature of the mirror $R$ is small.
More specifically, the fine structure is proportional to $\lambda/R$ and typically observable when $F \frac{\lambda}{R}>10$, where $F$ is the cavity finesse \cite{Siegman, Exter2022}.
The intrinsic corrections dominate over external effects, such as astigmatism, when $\lambda/R$ is large enough.

The frequency splittings that have been reported in literature are typically for cavities with larger radii of curvature, where the external effects of astigmatism \cite{Benedikter2015, Benedikter2019, Uphoff2015, papageorge2016} and birefringence \cite{Fleisher2016} dominate.
Intrinsic effects of fine structure have been reported for microwave cavities \cite{Yu1983, Erickson1975}, where $\lambda$ is large.
To the best of our knowledge, the spin-orbit coupling for cavities has only been reported in the optical domain in a conference proceeding \cite{Zeppenfeld-CLEO}.
We observe and analyze the complete fine structure for optical microcavities with very small radii of curvature.

In this paper, we show how the intrinsic effects determine the fine structure. 
We scan the cavity length to obtain the spectra of four optical microcavities with radii of curvature between $R=2.5(5)-17.3(2)~\mu$m and present the full analysis for the $R=5.8(2)~\mu$m cavity.
We first label the resonant modes in the spectrum with transverse order $N$, according to paraxial theory.
We then observe fine structure by zooming in on each $N$ group.
Using a polarization-resolved CCD camera, we can further identify every mode in the fine structure.
We study the fine structure systematically and compare it to theory.
Finally, we report a third type of splitting, which we call \enquote{hyperfine splitting} and which is due to a Bragg effect \cite{Foster2009}.

\begin{figure}
    \centering
    \includegraphics[width=\linewidth]{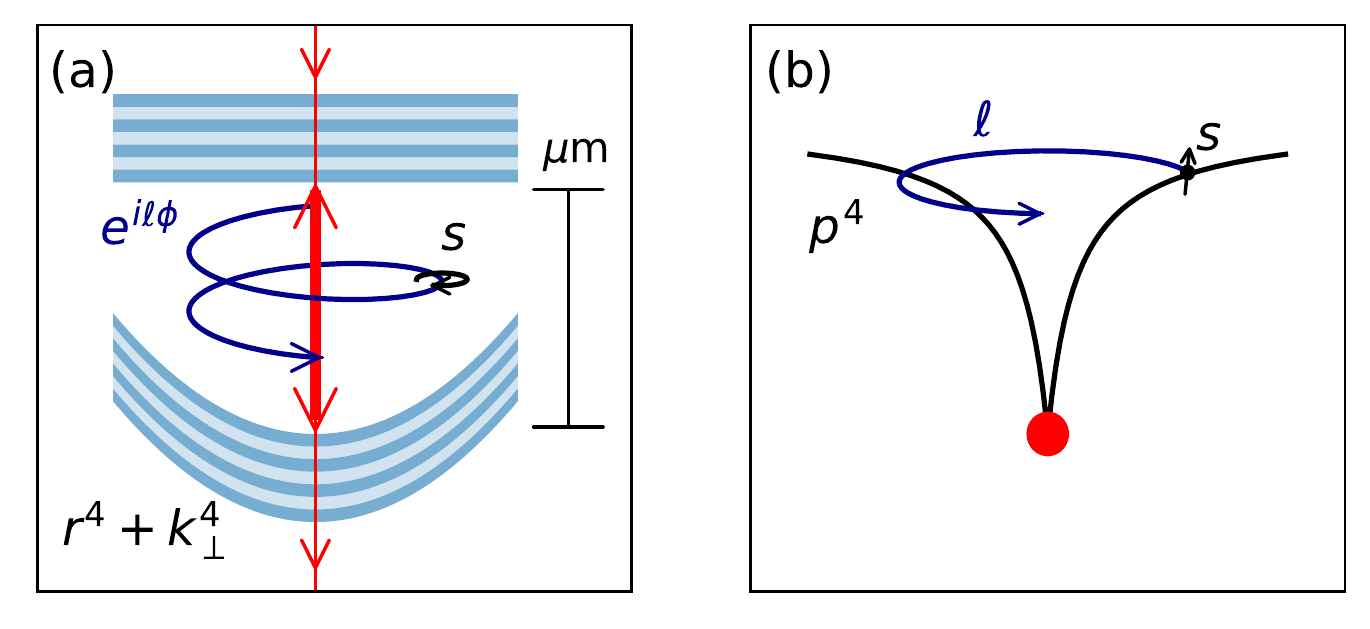}
    \caption{Analogy between optical microcavity and atom. (a) Optical field with spiral wavefront $e^{i l \phi}$ and circular polarization $s$ between a flat and curved mirror. (b) Electron wavefunction with orbital angular momentum $\ell$ and electron spin $s$.}
    \label{fig:introduction figure}
\end{figure}

\begin{figure*}
    \centering
    \includegraphics[width=\textwidth]{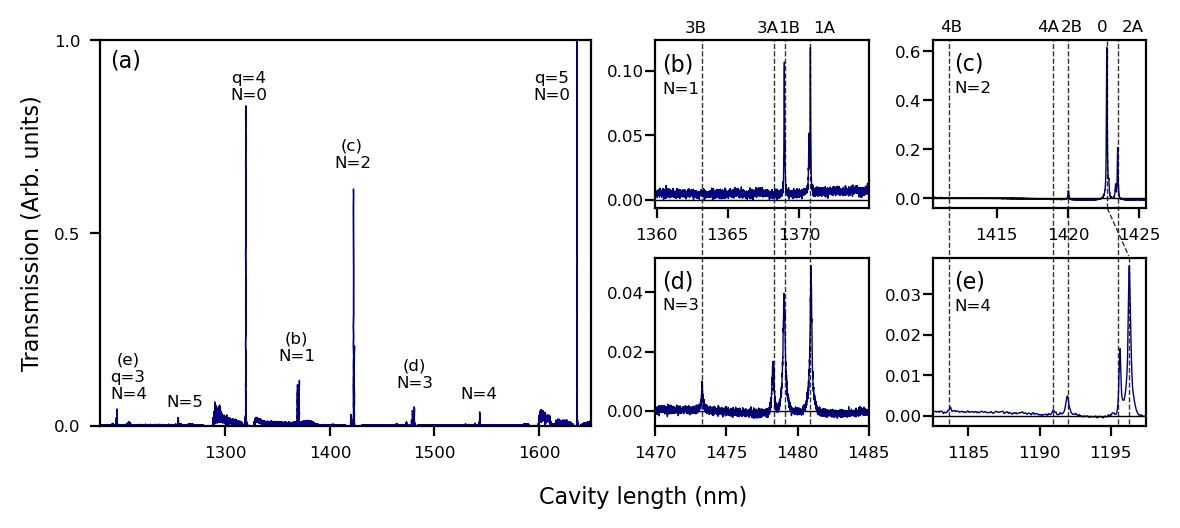}
    \caption{Cavity transmission spectrum shows fundamental and higher-order transverse modes. (a) Transmission versus cavity length as a function of the cavity length, where $q, N$ indicate the longitudinal and transverse mode number. (b-e) Zoom-ins of the groups N=1 to N=4 show fine structure.
    The broad peaks around $L=1300$ nm and $L=1600$ nm are resonances of planar modes formed next to the microcavity. 
    }
    \label{fig:full transverse mode spectrum}
\end{figure*}


The origin of the microcavity fine structure is similar to that of atoms.
Figure \ref{fig:introduction figure} illustrates the perturbations in both systems.
In atoms, spin-orbit coupling couples the orbital angular momentum $\ell$ and the spin $s$ of an electron through the magnetic field.
In microcavities, spin-orbit coupling couples the angular momentum $\ell$ and circular polarization spin $s$ of light \cite{Bliokh2015}.
The optical spin-orbit coupling originates from a projection of the longitudinal component of the electric field \cite{Yu1984} into an additional small transverse component at the mirror surface.
The relativistic correction in atoms is a quartic $p^4$ correction to the momentum, which shifts all modes proportional to $\ell^2$. As a direct analogy, a non-paraxial momentum correction $k_\perp^4$ is required for microcavities with large opening angles \cite{Lax1975, Erickson1975}, which also shifts all modes proportional to $\ell^2$.
In addition, the non-paraxial theory contains a $r^4$ correction from higher-order Taylor expansions of the mirror and wavefront shape \cite{Luk1986}.
The complete theory of microcavity fine structure is presented in \cite{Exter2022}.

Paraxial theory predicts resonant cavity lengths that only depend on the longitudinal mode number $q$, transverse order $N$, and Gouy phase $\chi=\arcsin\sqrt{(L+2L_{\rm D})/R}$, where $L_{\rm D}$ is the modal DBR penetration depth \cite{Koks2021}.
We experimentally determine the radius of curvature from the transverse mode spacings between each N group, which are equidistant in the paraxial theory \cite{Koks2021,Koks2022}.
A more complete (non-paraxial) description from \cite{Exter2022} contains the fine structure splittings $\Delta\tilde{L}$, 
\begin{equation}
    L=\frac{\lambda}{2}\left[q+\frac{N+1}{\pi}\chi+ \Delta\tilde{L}\right],
    \label{eq:paraxial solution}
\end{equation}
where
\begin{equation}
    \Delta\tilde{L}=\frac{1}{2\pi k R}\left[-\ell\cdot s-\left(\frac{3}{8}-\tilde{p} \frac{L}{8(R-L)}\right)\ell^2+f(N)\right].
    \label{eq: finestructure}
\end{equation}
This equation includes the two corrections: (i) the spin-orbit coupling, scaling with $l\cdot s$, and (ii) the quartic corrections $k_\perp^4$ and $r^4$, scaling with $\ell^2$.
The quartic corrections shift the modes by a factor $3 \ell^2/8$ when using a perfectly spherical mirror.
Perturbations to this mirror shape are quantified by the aspheric correction $\tilde{p}$ defined as $z_{\rm mirror}-z_{\rm sphere}=-\tilde{p} \frac{r^4}{8 R^3}$.
The term $f(N)$ shifts all modes of transverse order $N$ by the same amount and goes unnoticed in the fine structure.

Our planar and curved Distributed Bragg Reflectors (DBR) are produced by Oxford HighQ \cite{Trichet2015} and have a reflectively of 99.9\% (finesse $F\approx3000$).
The curved mirror is illuminated with a HeNe laser ($\lambda=633$ nm).
The cavity length is scanned with piezo-stacks and the light is transmitted through the microcavity at resonant cavity lengths.
This transmitted light is detected with a photodiode and a polarization-resolving CCD camera.

\begin{figure*}
    \centering
    \includegraphics[width=\textwidth]{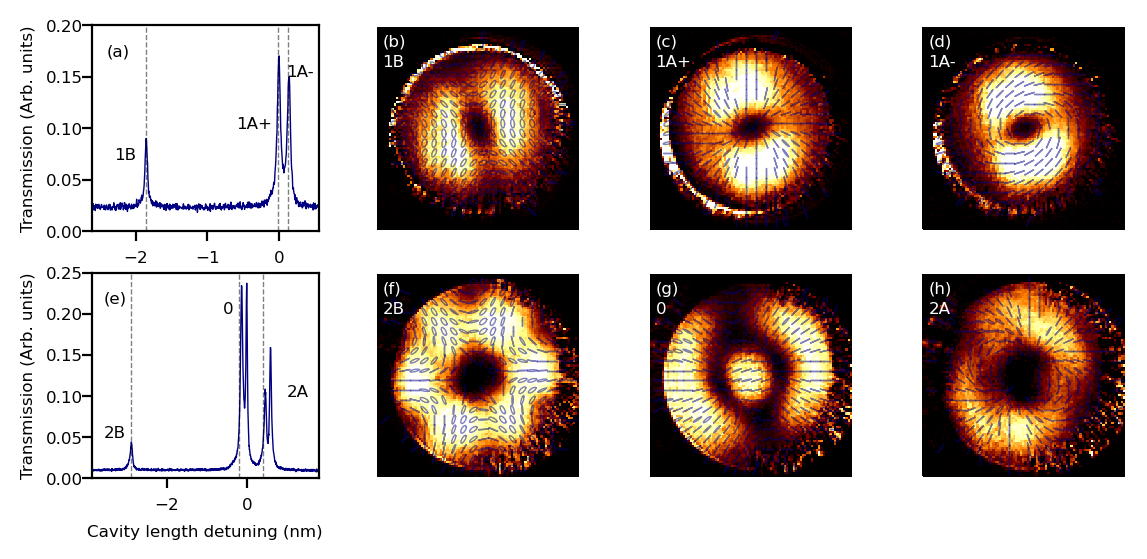}
    \caption{Fine structure splittings of the (a-d) N=1 and (e-h) N=2 transverse mode groups. The dashed lines in (a) and (e) correspond to polarization resolved CCD images (b-d) and (f-h), where the order from left to right corresponds to a smaller to larger cavity length detuning in (a) and (e). Each ellipse shows the local direction and circularity of the polarization.
    }
    \label{fig:splitting and modes}
\end{figure*}

Figure \ref{fig:full transverse mode spectrum}a shows a typical microcavity transmission spectrum, for the $R=5.8(2)~\mu$m cavity.
The peaks are located at resonant cavity lengths $L$. 
We can label them with $q$ and $N$ according to paraxial theory, which predicts that each transverse group $N$ consists of $2(N+1)$ orthogonal modes.
Figures \ref{fig:full transverse mode spectrum}b-e zoom in on each $N$ group.
This shows that, in practice, each group typically consists of $N+1$ modes.
The dashed lines suggest that all odd (or even) $N$-groups have similar mode and fine structures, albeit that larger $N$-groups contain more modes.
The $0$-mode for the even $N$ groups is shifted due to the shape of the mirror (see below).

Figure \ref{fig:splitting and modes} shows the spectrum and CCD images of the polarized eigenmodes of the $N=1$ and $N=2$ groups.
The eigenmodes of the $N=3$ and $N=4$ groups are shown in the Supplemental Document.
The mode labels $\ell A$ and $\ell B$ ($\ell>0$) in Fig. \ref{fig:full transverse mode spectrum} and \ref{fig:splitting and modes} are identified from the CCD images as follows.
The angular momentum $\ell$ is determined by inspecting the intensity profile and comparing it to the scalar Laguerre-Gaussian modes \cite{Siegman}.
For instance, the $N=2$ modes in Figs. \ref{fig:splitting and modes}f and \ref{fig:splitting and modes}h have a dark center and one ring, corresponding to $\ell=2$, whereas Fig. \ref{fig:splitting and modes}g has a bright center, corresponding to $\ell=0$.
The $A/B$ labels are determined from the polarization patterns, where the pattern of the $A/B$ modes resemble circular/hyperbolic flow lines. 
The total angular momentum $j=l+s$ ($s=-1$ for $A$ and $s=1$ for $B$ modes) is visible in the rotation symmetry of the polarization pattern, which remains unchanged after rotation over an angle $\pi/j$ (rotational symmetric at $j=0$).

Figure \ref{fig:splitting and modes}a also shows \enquote{hyperfine} splitting for mode $1A$.
This supports the theoretically prediction that each of the $N+1$ modes is typically two-fold degenerate and that this degeneracy can be lifted under certain conditions.
Those modes are labeled $+/-$ and have orthogonal polarization patterns.
The CCD images of the modes $1A-$ and $1A+$, shown in Figs. \ref{fig:splitting and modes}c and d, show that their polarization is in the azimuthal and radial direction, respectively.
Figure \ref{fig:splitting and modes}e also shows a hyperfine splitting for the $0$ and $2A$ modes.
Fig. \ref{fig:splitting and modes}g and \ref{fig:splitting and modes}h show the modes with the mostly radial polarization direction, which correspond to the left peaks in the hyperfine splitting of the $0$ and the $2A$ modes in Fig. \ref{fig:splitting and modes}e.

\begin{figure}
    \centering
    \includegraphics[width=\linewidth]{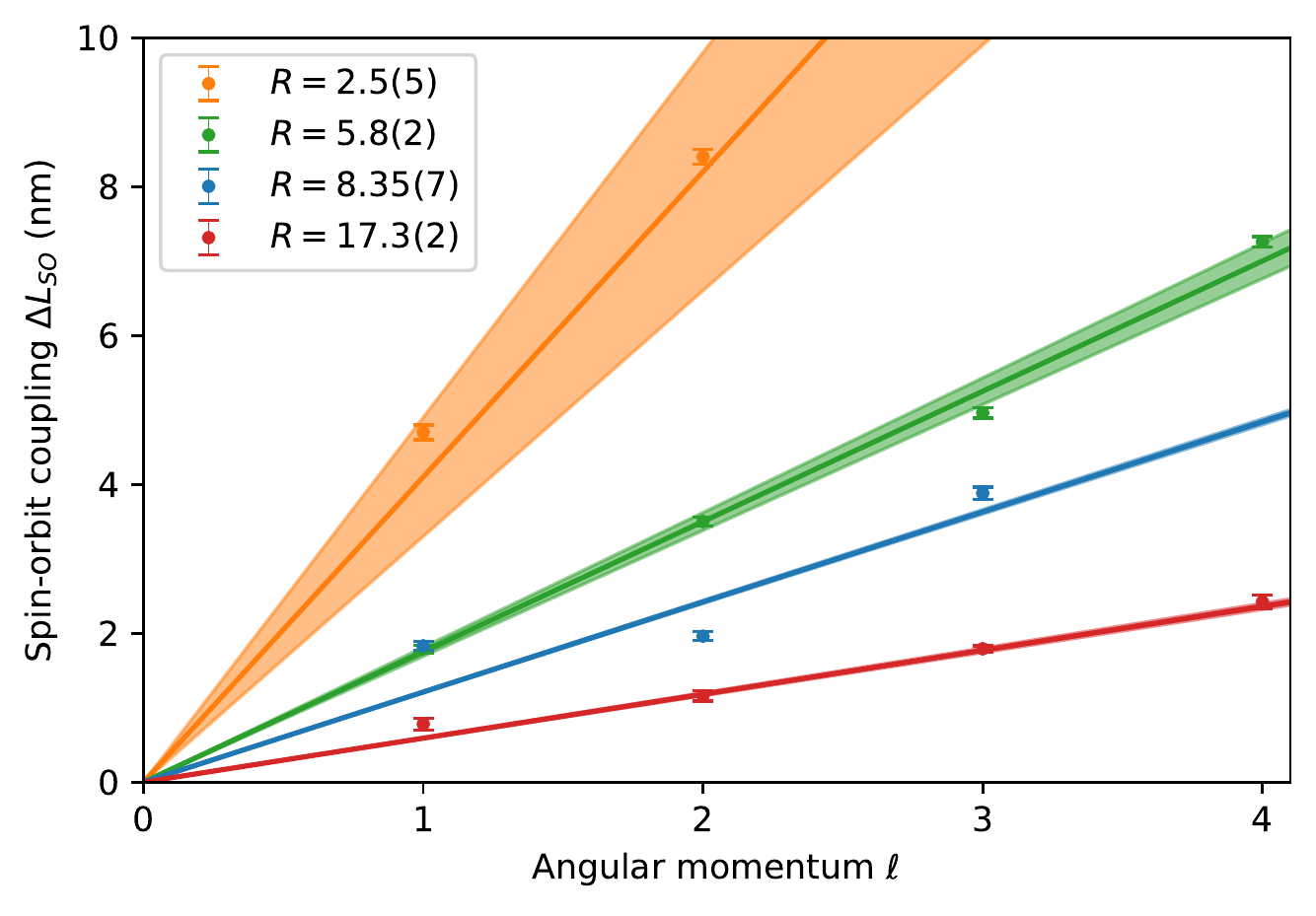}
    \caption{Observed mode splitting due to spin-orbit coupling for four different cavities. The lines show the theoretically predicted value with uncertainty for each radius of curvature.}
    \label{fig:spinorbitcoupling}
\end{figure}

Fig. \ref{fig:spinorbitcoupling} shows the measured splitting $\Delta L_{SO}$ between the $\ell A$ and $\ell B$ modes due to spin-orbit coupling.
The green points correspond $R=5.8(2)~\mu$m cavity presented in Figs. \ref{fig:full transverse mode spectrum} and \ref{fig:splitting and modes}, and scale linearly with the angular momentum $\ell$.
We observed the fine structure in three other cavities, which also show a clear linear scaling $\ell$.

Fig. \ref{fig:spinorbitcoupling} also shows the theoretical prediction of spin-orbit coupling $\Delta L_{SO}$ based on the measured radius of curvature $R$. 
The figure shows that the measured splittings follow the theory well for all four cavities, showing that spin-orbit coupling in these cavities dominates over external perturbations.
It also shows the inverse proportionality with $R$ of the fine structure splittings.

Theory predicts that a quartic perturbation shifts both $\ell A$ and $\ell B$ modes by the same amount, such that their resonant cavity lengths decrease proportional to $\ell^2$.
The data in figure \ref{fig:full transverse mode spectrum} agrees reasonably well with this prediction.
To quantify this effect, we look at the average position of the $\ell A$ and $\ell B$ modes, given by $\ell\overline{AB}$, and compare it with $(\ell+2)\overline{AB}$.
From equation \ref{eq: finestructure} we find that such splittings are $\Delta L_{\rm quartic,th}/(\lambda/2)=3 (\ell+1)/(4 \pi k R)$ where $\ell$ is the lowest angular momentum of the modes.
Theory predicts for $N=\ell+2=2, 3, 4$ that $\Delta L_{\rm quartic,th}=1.31(5)$ nm, $2.62(9)$ nm and $3.9(1)$ nm.
The measured splittings are $\Delta L_{\rm quartic}=0.93(5)$ nm, $4.17(7)$ nm and $6.3(1)$ nm. 
The measured values have the same sign and order of magnitude as the theoretical values but differ because of an aspheric correction $\tilde{p}\frac{L}{8(R-L)}=0.11(1),~ -0.20(1)$,  and $ -0.23(1)$.
The decreasing value for $\tilde{p}$ suggests that the cavity is flatter for compact (low $N$) modes and steeper for larger (high $N$) modes.
This agrees with the \enquote{bathtub} shape which has previously been observed in AFM measurements \cite{Koks2022}.

The hyperfine splitting of the $1A$ modes in Fig \ref{fig:splitting and modes}a can be explained by the Bragg effect.
It occurs because the DBRs have an angle-dependent penetration depth, which is opposite for radial ($1A+$) and azimuthal ($1A-$) polarized light \cite{Foster2009, Babic1993}.
The measured distance of $0.12(2)$ nm between the $1A+$ and $1A-$ modes can be explained by a small wavelength detuning from the stopband center of the DBR.
The hyperfine splitting of 0.15(2) nm of the $0$ and $2A$ modes in Fig. \ref{fig:splitting and modes}a can also be explained by the Bragg effect.
The $0$ and $2A$ modes mix due to astigmatism, such that the mode profile are more radially and azimuthally polarized. 
The mixing ratio of the $0$ and $2A$ modes is almost the same, which explain why the hyperfine splitting is also almost the same (see supplemental material).

Finally, we model the effect of astigmatism on the fine structure splitting by combining the intrinsic effects with aspherical and astigmatic corrections (see Supplemental Document).
When we fit this extended model to the splittings of the fine structure for $N=4$ in Fig. \ref{fig:full transverse mode spectrum}e, we find that the astigmatism $\eta_{\rm astig}=(R_{\rm max}-R_{\rm min})/(2 R)\approx0.6(2)\%$ for our $R=5.8(2)~\mu$m cavity is moderate.
Theory predicts astigmatism to be important when $\eta_{\rm astig}>1/(k\sqrt{L R})$, but the relative strength of astigmatic coupling between modes also depends on their frequency difference \cite{Koks2022}; modes with large angular momentum $\ell$ have a larger splitting and are therefore less sensitive to astigmatism.

In conclusion, we have observed fine structures in the resonance spectra of microcavities.
The fine structure is explained by two intrinsic perturbations.
First, the spin-orbit coupling causes a frequency splitting between the $\ell A$ and $\ell B$ modes, which scales with $\ell$.
Second, the quartic terms $k_\perp^4$ and $r^4$ shift both $\ell A$ and $\ell B$ by the same amount, scaling with $\ell^2$.
A parameter $\tilde{p}$ was introduced to quantify the aspheric shape of the mirror.
Furthermore, a hyperfine splitting was observed in the $1A+/1A-$ mode, which could be explained by the polarization dependence of the penetration depth in the DBR.
Measurements on other cavities with different radii of curvature showed that the fine structure splittings scale with $\lambda/R$.
The analogy with fine structure in atomic physics helps to understand the full spectrum and even some aspects of the hyperfine structure.

The efforts in the past years to increase light-matter interaction \cite{Wang2019, Najer2019} have simultaneously promoted an increase in finesse and decrease in radius of curvature.
The effects of fine structure are relevant when $F \frac{\lambda}{R}>10$, so fine structure will be large and unavoidable in such microcavities.
Moreover, if $\lambda/R$ is small, the intrinsic effects that cause fine structure tend to dominate over external effects such as astigmatism.
For a small radius of curvature, spin-orbit coupling, the quartic term, and the Bragg effect are essential to fully describe the mode profiles.

We would like to thank A. A. P. Trichet from Oxford HighQ for providing us with the mirror samples. We also acknowledge Sean van der Meer and Martin Bijl for supporting experiments, Martijn Wubs for supporting theory and Pepijn Pinkse for stimulating discussions.

\bibliographystyle{apsrev4-1}
\bibliography{main}

\end{document}


\title{Supplemental document: Observation of microcavity fine structure}
\date{\today}

\author{C. Koks}
\author{F.B. Baalbergen}
\author{M. P. van Exter}
\affiliation{Huygens-Kamerlingh Onnes Laboratory, Leiden University,
P.O. Box 9504, 2300 RA Leiden, The Netherlands}

\begin{abstract}
In this document, we present details on the measurements and analysis presented in the main article.
First, we show the polarization resolved CCD images for the $N=3$ and $N=4$ modes and determine their proper labels.
Second, we present a model that includes the effects of astigmatic and aspheric corrections. 
Last, we consider two effects that can cause hyperfine splitting of the $0$ and $2A$ modes of the $N=2$ group.
\end{abstract}

\maketitle

\section{Fine structure of the $N=3$ and $N=4$ modes}

\begin{figure}[H]
    \centering
    \includegraphics[width=\linewidth]{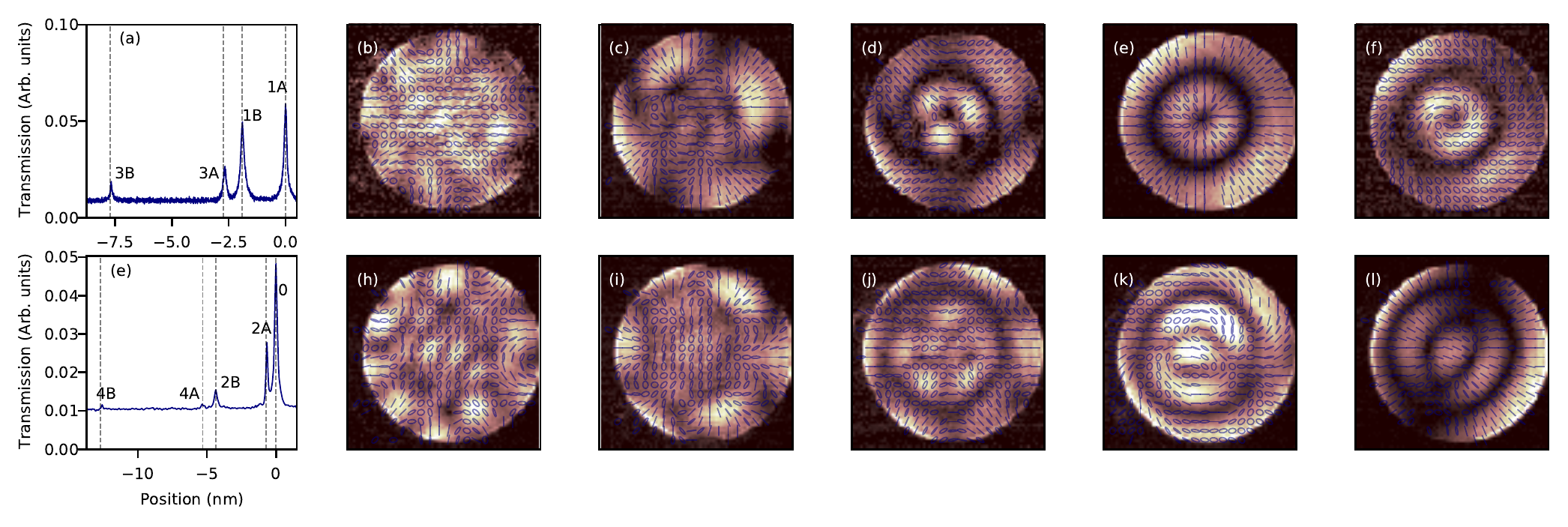}
    \caption{
    Fine structure splittings of the (a-f) N=3 and (g-i) N=4 transverse mode groups. The dashed lines in (a) and (g) correspond to the polarization resolved CCD images (b-f) and (h-l), where the order from left to right corresponds to a smaller to larger cavity length detuning in (a) and (g) and where (e) and (f) are two images of the $1A$ mode. Each ellipse shows the local direction and circularity of the polarization.
    }
    \label{fig:N=34 modes}
\end{figure}

Figure \ref{fig:N=34 modes} shows the polarization resolved CCD images of the $N=3$ and $N=4$ modes for the $R=5.8(2)~\mu$m cavity. 
The modes are labeled with $\ell A$ and $\ell B$ by inspection of their CCD images, similar to the labeling of the $N=1$ and $N=2$ group in the main article.
This labeling is however more subtle as we explain below.

The CCD images in Figs. \ref{fig:N=34 modes}e and \ref{fig:N=34 modes}f for $N=3$ are both labeled with $1A$.
Unlike the $1A+/-$ modes for $N=1$, we do not observe a clear spectral splitting between these modes.
We can still image both modes by rotating the polarization of the incoupling laser.
The input beam was injected slightly off-axis, such that a horizontal input polarization mostly exited the $1A-$ mode in figure \ref{fig:N=34 modes}f and a vertical input polarization the $1A+$ modes in figure \ref{fig:N=34 modes}e.

The $3B$ ($N=3$) mode in Fig. \ref{fig:N=34 modes}b and $4B$ ($N=4$) mode in Fig. \ref{fig:N=34 modes}f are difficult to identify because their images are limited by the numerical aperture (\textit{NA}$=0.5$) of the collection optics.
But polarization profiles at the edges are clear enough to determine the $A$ or $B$ label.
The modes also show a bright center, which is most likely due to the admixture of planar or higher-order modes.

The modes in Figs. \ref{fig:N=34 modes}k and \ref{fig:N=34 modes}l should be labeled with $0$ and $2A$, but both CCD images show a bright center, which corresponds to the $0$ mode. 
This shows the presence of mixing of the $2A$ and the $0$ modes, where observed modes are both partially $0$ and $2A$.
The next section presents a model which quantifies aspheric and astigmatic corrections and shows that they indeed cause this admixture.
We fit this model to the measured frequency splittings and find modes that look similar to the modes in Figs. \ref{fig:N=34 modes}k and \ref{fig:N=34 modes}l.
From this fit, we find that Fig. \ref{fig:N=34 modes}k is \enquote{mostly $2A$} and Fig. \ref{fig:N=34 modes}l is \enquote{mostly $0$} (see below).
We use this to label the modes in the resonance spectrum in Fig. \ref{fig:N=34 modes}g with $2A$ and $0$, respectively.

\section{Combined effect of spherical and astigmatic corrections}

We model the combined effect of the astigmatic and spherical corrections with two dimensionless parameters $X$ and $\tilde{p}$ that co-determine the $(N+1)\times(N+1)$ dynamic matrix.
The diagonal elements are the resonance frequencies of the unperturbed $\ell A$ and $\ell B$ modes which form the basis of this matrix.
The diagonal elements are co-determined by aspheric corrections, whereas the off-diagonal elements are determined by the astigmatic corrections. 
More elements of this model are described in \cite{Exter2022}.

The aspheric correction $\tilde{p}$ shifts the resonance frequencies by $2 \pi k R\Delta \tilde{L}=\tilde{p} \frac{L}{8(R-L)} l^2$, with radius of curvature $R$ and cavity length $L$.
The parameter $\tilde{p}$ is defined as $z_{mirror}-z_{sphere}=-\tilde{p} r^4/(8 R^3)$, where $\tilde{p}=0$ for a spherical mirror shape and $\tilde{p}=1$ for a parabolic mirror shape.
The astigmatic correction $\eta_{astig}$ couples the $\ell A/B$ modes with $(\ell\pm 2) A/B$ modes, to form new modes.
For only slightly ellipsoidal mirrors, we define $\eta_{astig}=(R_{max}-R_{min})/(2R)$, where $R_{max}$ and $R_{min}$ are the radii of curvature along the long and short axes of the mirror.
We also define a relative astigmatism $X=4 k R \eta_{astig} \tan(\chi)$, with Gouy phase $\sin\chi=\arcsin\sqrt{L/R}$, which is unity when astigmatic and intrinsic effects on the fine structure are approximately the same.

\begin{figure}[H]
    \centering
    \includegraphics[width=\textwidth]{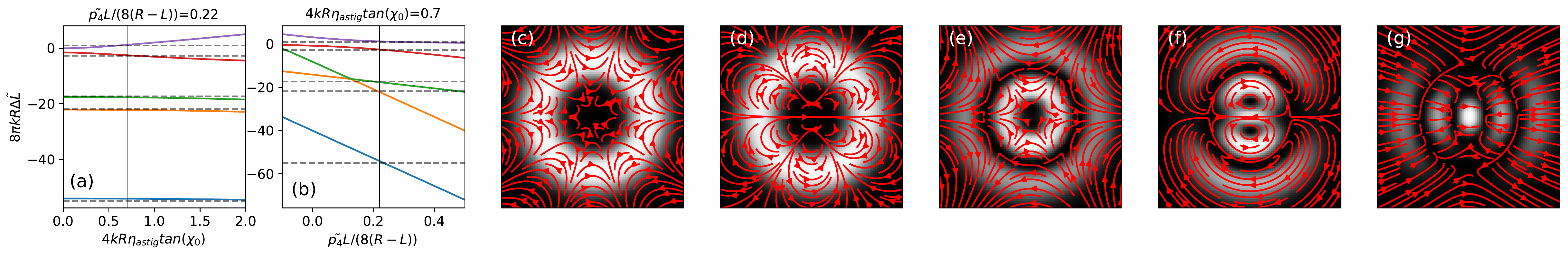}
    \caption{Simulation of the combined effects of astigmatic and aspheric correction on resonance frequencies and mode profiles. (a-b) Frequency shifts due to (a) astigmatic and (b) aspheric corrections around the fitted values (vertical lines) deduced from the measured frequencies (horizontal dashed lines) of the $N=4$ modes in the $R=5.8(2)~\mu$m cavity. (c-g) Mode profiles in order of frequency at the vertical lines in (a-b); mode in (c) has lowest frequency.
    }
    \label{fig:astigmatism and quartic therm model}
\end{figure}

We can fit this model to the measured fine structure spectrum of the $N=4$ group for the $R=5.8(2)~\mu$m cavity using two fit parameters.
This fit yields an aspheric correction $\tilde{p} \frac{L}{8(R-L)}=-0.22(2)$ and relative astigmatism $X=0.7(2)$, which corresponds to an absolute astigmatism of $\eta_{astig}=0.6(2)$\%.
The fitted aspheric correction also agrees well with the measured value in the main text of $\tilde{p}=-0.23(1)$ for the $N=4$ group, where we only considered four modes and neglected astigmatism.
The calculated frequencies are plotted in Figs. \ref{fig:astigmatism and quartic therm model}a and \ref{fig:astigmatism and quartic therm model}b, which show the dependence on astigmatic and aspheric corrections around the measured frequencies for $N=4$ (horizontal dashed lines).

Figures \ref{fig:astigmatism and quartic therm model}c-g show the calculated eigenmodes for these fitted parameters.
The $4B$, $4A$ and $2B$ modes in Figs \ref{fig:astigmatism and quartic therm model}c-e seem unaffected, but the modes in Figs. \ref{fig:astigmatism and quartic therm model}f and \ref{fig:astigmatism and quartic therm model}g show strong signs of mixing.
The calculated modes in Figs. \ref{fig:astigmatism and quartic therm model}c-g strongly resemble the measured modes in Figs. \ref{fig:N=34 modes}h-i.

The mixed modes in Figs. \ref{fig:astigmatism and quartic therm model}f and \ref{fig:astigmatism and quartic therm model}g can approximately be written as a superposition of the $0$ and $2A$ modes, $\ket{\psi}=\alpha\ket{2A}+\beta\ket{0}$, where $|\alpha|^2+|\beta|^2=1$.
The calculated eigenvectors that follow from the fit show that Fig. \ref{fig:astigmatism and quartic therm model}f is \enquote{mostly $2A$} ($|\alpha|>|\beta|$) and Fig. \ref{fig:astigmatism and quartic therm model}g is \enquote{mostly $0$} ($|\alpha|<|\beta|$).

In Fig. \ref{fig:astigmatism and quartic therm model}b, the purple and red line shows an avoided crossing around the fitted value (vertical line).
The red line is \enquote{mostly $0$} before the crossing when $-\tilde{p} \frac{L}{8(R-L)}< 0.1$ and \enquote{mostly $2A$} after the avoided crossing when $-\tilde{p} \frac{L}{8(R-L)}>0.3$, and the purple line is \enquote{mostly $2A$} before the crossing and \enquote{mostly $0$} after.
The fitted value in the vertical line is located slightly after the avoided crossing, making the frequency of the \enquote{mostly $0$} mode higher than that of the \enquote{mostly $2A$} mode. 
This explains the change in the relative locations of the $0$ and $2A$ modes for the $N=2$ and $N=4$ groups (main text).
We measured a positive value $\tilde{p}=0.11$ for the $N=2$ group, hence it is before the avoided crossing where the frequency of the \enquote{mostly $2A$} mode is higher than that of the \enquote{mostly $0$} mode.

\section{Hyperfine splitting}
The $0$ and $2A$ modes ($N=2$) in the main text show hyperfine splittings of 0.15(2) nm.
In this section, we estimate the contribution of astigmatism to hyperfine splitting. 
We consider two mechanisms by which astigmatism can cause hyperfine splitting: (i) the Bragg effect and (ii) shape birefringence.

\subsection{Bragg effect}

\begin{figure}[H]
    \centering
    \includegraphics[scale=0.7]{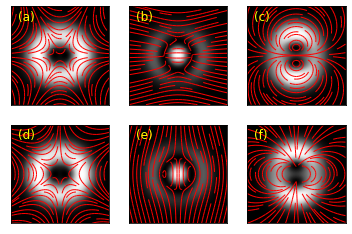}
    \caption{Calculated modes of the N=2 group: (a-c) modes of the \enquote{+} group, (d-f) modes of the \enquote{-} group. From left to right the 2B, 0 and 2A modes in order of increasing resonance frequencies. The red lines indicate the polarization profile.}
    \label{fig: N=2 +- bragg effect modes}
\end{figure}

The Bragg effect originates from a difference between the penetration depth of radial and azimuthal polarization.
We observed a hyperfine splitting of 0.12(2) nm for the $1A+$ and $1A-$ modes, which we attributed to the Bragg effect because the $1A\pm$ modes are purely radial/azimuthal polarized.
At first sight, one would expect that other modes are not affected by the Bragg effect, because their polarization patterns contain equal amounts of radial and azimuthal polarization and rotational symmetry averages out the penetration depths.
However, astigmatism breaks rotational symmetry, as the mixed eigenmodes for the $N=2$ group in Fig. \ref{fig: N=2 +- bragg effect modes} show.
The modes in Figs. \ref{fig: N=2 +- bragg effect modes}b and \ref{fig: N=2 +- bragg effect modes}f have most intensity in the radial polarization direction and the modes in Figs. \ref{fig: N=2 +- bragg effect modes}c and \ref{fig: N=2 +- bragg effect modes}e in the azimuthal polarization direction.
This modified intensity distribution causes the Bragg effect for the modes other than $1A$.

We calculate the hyperfine splitting of the \enquote{mostly 2A} mode in Figs. \ref{fig: N=2 +- bragg effect modes}c and \ref{fig: N=2 +- bragg effect modes}f and \enquote{mostly 0} mode in Figs. \ref{fig: N=2 +- bragg effect modes}b and \ref{fig: N=2 +- bragg effect modes}e relative to the hyperfine splitting of the $1A$ mode of the $N=1$ group (see also \cite{Exter2022}).
The modes are written as mixed modes of the form $\ket{\psi_{2A\pm}}=\alpha\ket{2A}\pm\beta\ket{0}$ and $\ket{\psi_{0\pm}}=\alpha'\ket{2A}\pm\beta'\ket{0}$ with $|\alpha|^2+|\beta|^2=1$, $|\alpha'|^2+|\beta'|^2=1$.
The $\pm$ correspond to the top and bottom row in Fig. \ref{fig: N=2 +- bragg effect modes}.
We then calculate the perturbation of the Bragg effect $H_{\rm Bragg}$ \cite{Exter2022} of the $0$ and $2A$ modes, relative to the $1A$ modes: $\bra{0}H_{\rm Bragg}\ket{2A}=\sqrt{2}\bra{1A}H_{\rm Bragg}\ket{1A}$ and $\bra{0} H_{\rm Bragg} \ket{0}=\bra{2A} H_{\rm Bragg} \ket{2A}=0$.
The calculated hyperfine splitting of the \enquote{mostly 2A} and \enquote{mostly 0} modes is $\Delta\tilde{\nu}=2\sqrt{2}\alpha \beta\Delta\tilde{\nu}_{1A}$.
The values of $\alpha$ and $\beta$ are calculated from the model described above using the fitted astigmatism of $\eta_{\rm astig}=0.6(2)\%$ and the measured aspheric correction of $\tilde{p} \frac{L}{8(R-L)}=0.11(1)$ for the $N=2$ group.
The calculated parameters are $\alpha=0.97(2)$ and $\beta=0.24(6)$ for the mode which is \enquote{mostly $2A$} and $\alpha=0.24(6)$ and $\beta=0.97(2)$ for the mode which is \enquote{mostly $0$}.
The calculated hyperfine splitting for both the \enquote{mostly $2A$} and \enquote{mostly $0$} mode relative to the $1A$ mode is thus $\Delta \tilde{\nu}/\Delta\tilde{\nu}_{1A}=0.6(2)$.

Using the measured hyperfine splitting of 0.12 nm for the $1A+/-$, we thus expect that the hyperfine splitting of the $0$ and $2A$ modes is $\Delta L=0.07(3)$ nm.
Therefore, the Bragg effect can partly explain the measured hyperfine splitting of $0.15(2)$ nm.

\subsection{Shape Birefringence}
The ellipsoidal mirror shape can also cause birefringence between modes polarized along the long and short axes.
This shape birefringence was observed before in the hyperfine splitting of the fundamental mode ($N=0$) \cite{Uphoff2015, Mader2015}.
The effect on modes with non-vanishing angular momentum is more subtle, and the proper perturbation that contributes to this effect is (see also \cite{Exter2022})
\begin{equation}
    H_{SB}=\frac{\eta_{astig}}{\pi k R} (\hat{x}x-\hat{y}y)(\partial_x\hat{x}+\partial_y \hat{y}).
\end{equation}
Calculating the coupling between the $+$ and $-$ modes gives
\begin{equation}
    \Delta \nu_{SB}= 2 \alpha \beta \bra{0\pm} \frac{H_{SB}}{2\pi} \ket{2A\pm}=\mp2 \alpha \beta \frac{\eta_{astig}}{4\pi k R}.
\end{equation}
For $\eta_{astig}=0.6(2)\%$ and $\tilde{p} \frac{L}{8(R-L)}=0.11(1)$, the calculated hyperfine splitting is $0.0012(6)$ nm and therefore we can safely neglect it. 
Hence, we conclude that the hyperfine splitting in this particular cavity is caused by the combined action of astigmatism and a Bragg effect and that shape birefringence plays a negligible role.

\subsection{Large astigmatism}
\begin{figure}[H]
    \centering
    \includegraphics{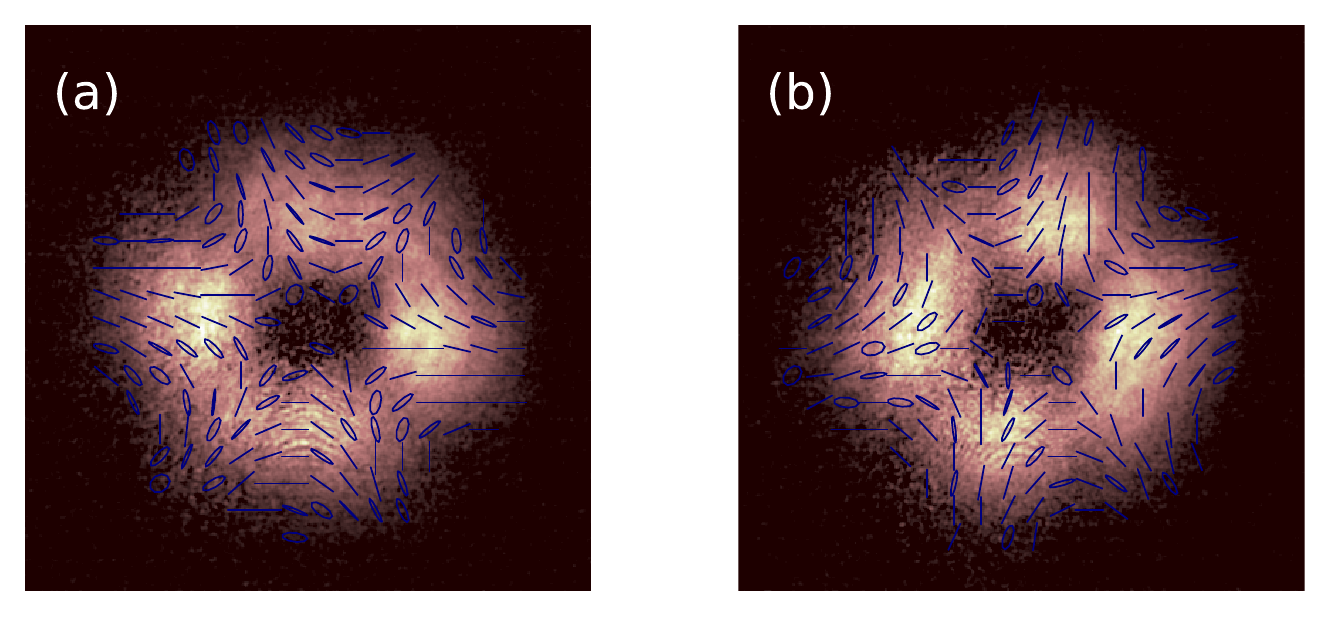}
    \caption{Measured 2B modes for the N=2 group in a $R=17.1~\mu$m cavity. (a) Shows the 2B+ mode and (b) the 2B- mode. Each ellipse shows the local direction and circularity of the polarization.}
    \label{fig:hyperfine 2B modes}
\end{figure}

The effect of shape birefringence can become more important in case of large astigmatism.
Fig. \ref{fig:hyperfine 2B modes} shows the $2B$ modes for a $R=17.1(2)~\mu$m cavity with 5\% astigmatism.
We observed a hyperfine splitting of 0.22(3) nm between the $2B+$ and $2B-$ modes. 
The polarization profiles of the eigenmodes show that most intensity is in the horizontal polarization direction for mode $2B+$ in Fig. \ref{fig:hyperfine 2B modes}a and in the vertical polarization direction in Fig. \ref{fig:hyperfine 2B modes}b.
Therefore, the Bragg effect should average out, since the modes contain equally as much radial as azimuthal polarization.
The shape birefringence is a more plausible cause for the splitting since this acts on the horizontal or vertical polarization.

\bibliographystyle{apsrev4-1}
\bibliography{main2}